\documentstyle[sprocl2,epsfig]{article}

\def\be{\begin{equation}}
\def\ee{\end{equation}}
\def\bea{\begin{eqnarray}}
\def\eea{\end{eqnarray}}

\begin{document}

\title{Two loop renormalization of the magnetic coupling in hot QCD}

\author{P. Giovannangeli }

\address{Centre Physique Th\'eorique au CNRS,  
Case 907, Luminy 
F13288, Marseille, France\\
E-mail: giovanna@cpt.univ-mrs.fr} 


\maketitle\abstracts{Well above the critical temperature hot QCD is described
by 3d electrostatic QCD with gauge coupling $g_E$ and Debye mass $m_E$. We integrate out the Debye scales to  two loop accuracy and find for the  gauge coupling in the resulting magnetostatic action $g_M^2=g_E^2(1-{1\over{48}}{g_E^2N\over{\pi m_E}}-{17\over{4608}}({g_E^2N\over{\pi m_E}})^2+O(({g_E^2N\over{\pi m_E}})^3))$.}

\section{Introduction}

Notable progress~\cite{kajantie} in the standard model at high $T$
is due to the systematic separation of perturbative scales like  the temperature, the Debye mass $gT$ and the non-perturbative scale $g^2T$. It is the latter
that is three dimensional and can be treated numerically on the lattice and has given us a wealth of information on the plasma state of the standard model and QCD itself. In this note we will be concerned with QCD, but we will admit for $N$ instead of three colours.

 For small gauge coupling $g$ one can integrate out the integer Kaluza-Klein modes $2\pi nT$ and obtain a static effective action, $S_E$, with a running coupling $g_E(T)$. This action is three dimensional, and its degrees of freedom are three dimensional Yang-Mills and a massive adjoint Higgs  with as mass the  Debye scale $gT$. The scale $g^2T$ appears as the coupling in the three dimensional Yang-Mills action. The Debye scale 
 can be integrated out when $g<<1$. We are then left with the magnetic action $S_{M}$. It is the three dimensional Yang-Mills theory which describes physics at the 
magnetic scales $g^2T$. Its coupling $g_M^2$ is a function of the parameters in
$S_E$ and can be calculated perturbatively. This has up to now been done to one loop order~\cite{shapfara}. In this letter we report on the computation of the two loop effects.

\section{Motivation}

Our motivation stems from the need for accuracy. More precisely, integrating out effects of the K-K modes one obtains from the original action of QCD the 
superrenormalizable action $S_E$:
\begin{eqnarray}
{\cal{L}}_{E} & = & Tr(\vec D(A)A_0)^2+m_E^2TrA_0^2+\lambda_E(Tr(A_0^2))^2+ {} \nonumber\\
              & + &  \bar\lambda_E \big(Tr(A_0)^4-{1\over 2}(TrA_0^2)^2\big)+{1\over 2}Tr F_{ij}^2+\delta{\cal{L}}_E.
\label{estat} 
\end{eqnarray} 

This action density describes the physics of the QCD plasma down to temperatures $T\sim 2T_c$, including non-perturbative effects from the magnetic density
$Tr F_{ij}^2$. These have been calculated by lattice methods~\cite{owe}. 
The terms neglected, $\delta{\cal{L}}_E$,  introduce an error of $O(g^4)$~\cite{shaposhkajantie}.
So the parameters in this action have all evaluated up to this order.

The magnetic action takes the form

\begin{equation}
{\cal{L}}_M={1\over 2}TrF_{ij}^2+\delta{\cal{L}}_M
\label{mstat} 
\end{equation}   
\noindent
with a magnetostatic gauge coupling $g^2_M$.

Now the  neglected terms introduce an error~\cite{shaposhkajantie} of $O(g^3)$, and 
hence the magnetic coupling has to be computed to $O(g^2)$ accuracy.
The calculation is reported on in the next  section.

\section{Renormalization of the magnetic gauge coupling}

The basic idea behind the effective actions eqns (\ref{estat}) and (\ref{mstat}) is that one can compute with both in the region of momenta $p\sim g^2 T$.
To know what the parameters of the latter are in terms of those of the former
 requires computing two-point functions, three point funtions etc.  in both theories and match them. In the matching the diagrams of the pure 3d Yang-Mills theory drop out.

Here we will follow a well-known shortcut~\cite{abbott} by introducing a background field 
$B_i$ in  ${\cal{L}}_{E}$:
\begin{eqnarray}
 \vec A & = &{1\over {g_E}}\vec B +  \vec Q\nonumber\\
    A_0 & = & g_EQ_0.
\end{eqnarray}
We calculate the fluctuations around the background in a path integral:
\begin{equation}
exp{-{1\over{g_M^2}}S_M(B)}=\int DQ_0DQ_i\exp{\big(-S_E-{1\over{\xi}}Tr(D_iQ_i)^2\big)}.
\label{path}
\end{equation}

\begin{figure}
\begin{center}
\includegraphics{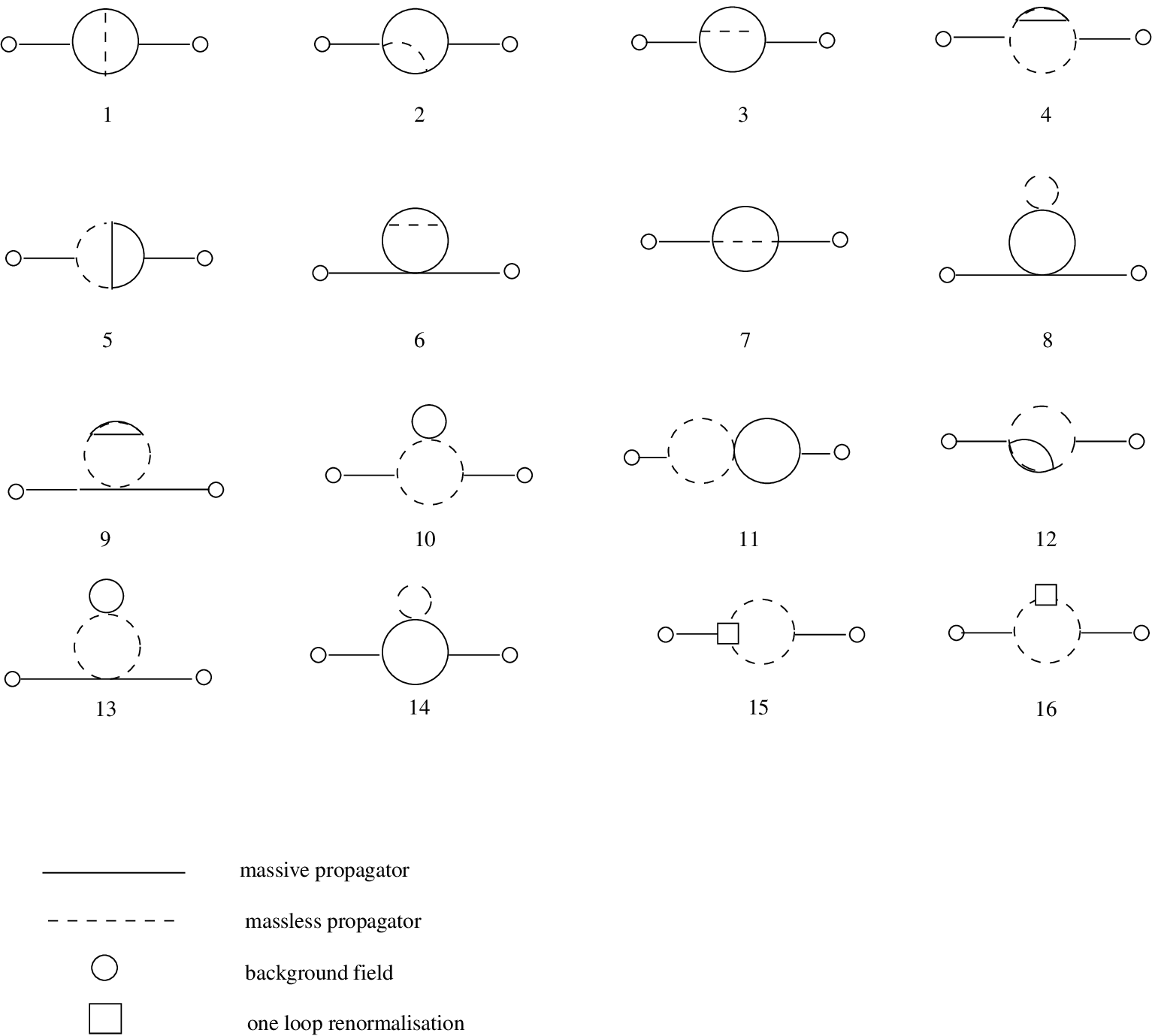}
\caption{two-loop Feynman graphs for the background two-point function}
\label{fig:graph}
\end{center}
\end{figure}

We added a general background gauge term. The resulting action $S_M(B)$ is 
gauge invariant to all loop orders and the renormalization of the coupling
is identified from the background field two point function at a momentum $p=O(g^2T)$. This momentum is the infrared cut-off in computing the 
r.h.s. of eq.(\ref{path}). With dimensional regularization one finds in d dimensions, dropping the pure Yang-Mills diagrams as mentioned before:
\begin{equation}
 exp{\big(-{1\over{g_M^2}}S_M(B)\big)}= exp{\big(-{1\over{g_E^2}}S_M(B)\big)}\big(1+(F_1^{tr}+F_2^{tr}+...)S_M(B)\big).
\label{ident}
\end{equation}
Here the $F_i^{tr}$ are the transverse parts of the background two-point functions $F_i$ as shown for the two loop case in fig.(\ref{fig:graph}).

 Let the sum of all Feynman diagrams  for the two point function of the background field with $i$ loops    be $F_i$. Then we can write:
\begin{equation}
F_i=F_i^{tr}~(\delta_{lm}p^2-p_lp_m)+F_i^L~p_lp_m
\end{equation}

The longitudinal part is zero because of gauge invariance. It is borne out by explicit calculation. Transversality is true for all $d$ and values of the parameters.

The $F_i^{tr}$ are still depending on the parameters $\epsilon = {3-d\over2}$, $g_E$, $m_E$, $\xi$, the regularization scale $\bar\mu$ and the momentum $p$. We are interested in the limits $d=3$ and
$\tilde{p}= 0$ where $\tilde{p} = {p\over m_E}$. In that limit we expect because of the superrenormalizabilty the UV poles and the $\bar\mu$ dependence to cancel. Also the $\xi$  dependence should disappear.  And so should the IR effects in the guise  of inverse powers of the momentum $p$.  And indeed they do by explicit calculation, as the table in the next section shows.

This leaves us with the relation, using eq.(\ref{ident}):
\begin{equation}
{1\over{g_M^2}}={1\over{g_E^2}}-F_1^{tr}-F_2^{tr},
\end{equation}

with
 
\begin{eqnarray}
g_E^2F_1^{tr} & = & -{1\over{48}}{g_E^2N\over{\pi m_E}}\\
g_E^2F_2^{tr} & = & -{19\over{4608}}({g_E^2N\over{\pi m_E}})^2.
\label{final}
\end{eqnarray}

This is the main result.
\section{Details of the calculation}

The two loop diagrams involving at least one massive propagator are 
shown in fig.(\ref{fig:graph}). Also shown (graphs 15 and 16) are the insertions, discussed in ref.(\cite{abbott}). They are vital for the gauge parameter independence of our result.
FORM~\cite{form} was used for algebraic manipulations and scalarisation of integrals with reducible numerators. The program TARCER~\cite{tarcer} served to scalarize those integrals with irreducible denominators. At that point the result $F_2$ is expressed in terms of eight scalar integrals. For $d=3$ , the values of these scalar integrals are computed in ref.(\cite{rajantie}). For the expansion of these integrals up to $\epsilon$, methods as in ref.(\cite{rajantie}) were used and results were checked with ref.(\cite{york}).

First we checked the transversality, i.e. $F_2^L=0$. The reader can find the result for $F_2^{tr}$ in table below. Individual graphs have UV and IR divergencencies that do cancel when summed. The finite part is indeed gauge parameter independent, though the physically irrelevant $O(p^2)$ terms are not.\\

\begin{tabular}{|c|c|}
\hline
\raisebox{0pt}[14pt][6pt]{Graph no} &
\raisebox{0pt}[14pt][6pt]{$p^2 F_2^{tr}$} \\
\hline 
&\\
1& $-{\tilde{p}^2(29+24\xi)\over9216\pi^2}+{1+\xi+\epsilon(2+\xi-4(1+\xi)\log(2))+4\epsilon(1+\xi)\log({\bar{\mu}\over m})\over128\epsilon\pi^2}$ \\
2& ${\tilde{p}^2(20+6\xi)\over1536\pi^2}-{3(2+\xi-2\epsilon(-1+\xi(-1+\log(4))+\log(16))+4\epsilon(2+\xi)\log({\bar{\mu}\over m})))\over128\epsilon\pi^2}$ \\
3& ${\tilde{p}^2(-43+12\xi)\over4608\pi^2}+{4+\epsilon(1+\xi-16\log(2))+16\epsilon\log({\bar{\mu}\over m})\over64\epsilon\pi^2}$ \\
4& ${-2-\xi\over32\tilde{p}\pi}-{\tilde{p}(15+\xi)\over1536\pi}+{\tilde{p}^2(13+2\xi)\over3072\pi^2}-{3(1+\xi)^2+4\epsilon(-8+\xi)\xi\log({\bar{\mu}\over2m})+4\epsilon(3+2\xi(7+\xi))\log({\bar{\mu}\over2m})\over768\epsilon\pi^2\xi}$ \\
5& $ {\tilde{p}(15+\xi(4+\xi))\over1536\pi}-{\tilde{p}^2(23+12\xi)\over4608\pi^2}+{2\epsilon+\xi+4\epsilon\xi\log({\bar{\mu}\over2m})\over128\epsilon\pi^2}$ \\
6& ${-3\over(64e\pi^2)}+{22-24(1+\log({\bar{\mu}\over2m}))\over128\pi^2}$ \\
7& ${6(2+\xi)+\epsilon(12(1+\xi)-\tilde{p}^2(2+\xi))+24\epsilon(2+\xi)
\log({\bar{\mu}\over2m})\over512\epsilon\pi^2}$\\
8& $0$\\
9& ${1+2\xi\over256\epsilon\pi^2\xi}+{-2\xi-4(1+2\xi)-4(-1-2\xi)(1+\log({\bar{\mu}\over2m}))\over256\pi^2\xi}$\\
10& ${16+8\xi\over256\tilde{p}\pi}$\\
11& $0$\\
12& $0$\\
13& $0$\\
14& $0$\\
15& $-{\tilde{p}(1+\xi)(4+\xi)\over1536\pi}$\\
16& ${\tilde{p}(2+\xi)\over768\pi}$\\
&\\
\hline
\hline
&\\
sum& ${-19\tilde{p}^2\over4608\pi^2}$\\
&\\
\hline
\end{tabular}

\section{Conclusions}

Our main result, eq.(\ref{final}), shows that the smallness of the corrections to the magnetic coupling does persist in two loop order. In fact, at $2T_c$ the coupling for 3 colours equals~\cite{owe}  $g^2_E=2.7$ and the 2 loop correction is about a third of the one loop correction (itself about 3 percent).

Our result is of importance in analyzing the  purely magnetic quantities,
as the spatial Wilson loop, and the magnetic mass. In particular it is crucial in connecting the lattice results from the magnetic action  to those obtained 
from the electric action, and ultimately to those of four dimensional simulations. This will be done in a future publication\cite{giovanna}. 
I thank Chris Korthals Altes for his help throughout my work.  
I acknowledge the help of Mikko Laine and York Schroder for very useful and stimulating advice.
I thank the MENESR for financial support.

\end{document}